\documentclass[11pt]{article}

\usepackage[margin=1in]{geometry}
\usepackage{amsmath,amssymb,amsthm}
\usepackage{algorithm}
\usepackage{algpseudocode}
\usepackage{graphicx}
\usepackage{hyperref}
\usepackage{xcolor}
\usepackage{booktabs}
\usepackage{tikz}
\usetikzlibrary{arrows.meta,positioning}

\newtheorem{theorem}{Theorem}

\newtheorem{conjecture}[theorem]{Conjecture}

\title{\textbf{On a Conjecture for Parameterized $st$-Orientations}}

\author{Charalampos Papamanthou\thanks{\texttt{charalampos.papamanthou@yale.edu}}\\ \emph{Yale University} }

\date{\today}

\begin{document}

\maketitle

\begin{abstract}
 
 \textsc{MaxSTN} and \textsc{MinSTN}---proposed by Papamanthou and Tollis~\cite{PapamanthouTollis2008,PapamanthouT10}---are two algorithms for
producing $st$-orientations of biconnected graphs with long and short longest paths respectively. Based on extensive experiments on planar and non-planar graphs of up to 5,000 nodes, it was conjectured (Conjecture 3.4.2~\cite{Papamanthou05}) that $\ell_{\max} \geq \ell_{\min}$ for every biconnected graph $G$, where $\ell_{\max}$ and $\ell_{\min}$ denote the longest-path
lengths of the two orientations. This paper disproves this conjecture by exhibiting a
biconnected graph on 9 vertices for which 
\textsc{MaxSTN} yields $\ell_{\max}=6$ while \textsc{MinSTN} yields $\ell_{\min}=7$, regardless of how ties are broken in
either algorithm. %The counterexample is tight: exhaustive search confirms
%the conjecture holds for all biconnected graphs on at most six vertices.
\end{abstract}

%-----------------------------------------------------------------------
\section{$st$-Orientations}
%-----------------------------------------------------------------------

An \emph{st-orientation} of a biconnected undirected graph $G=(V,E)$ is
an acyclic orientation of its edges such that exactly one vertex $s$
(the source) has in-degree zero and exactly one vertex $t$
(the sink) has out-degree zero~\cite{EvenTarjan76,Tarjan86}. A biconnected graph may admit many distinct
$st$-orientations.  Among all such orientations the length of the longest
directed path from $s$ to $t$ could vary, and  has important
implications in graph drawing and combinatorics.  For example, the longest-path length determines the height of any
    visibility representation of a planar graph~\cite{TamassiaT86} as well as the minimum number of layers required in certain upward planar drawings~\cite{DiBattistaETT99,KantHe97}.
Computing the $st$-orientation that maximizes the longest directed path
is NP-hard~\cite{PapamanthouTollis2008}, as is computing the one that
minimizes it.  This motivates polynomial-time algorithms that produce long or short longest paths respectively.

Papamanthou and Tollis~\cite{PapamanthouTollis2008} proposed a family of
algorithms for computing $st$-orientations based on a technique they call STN.  Two members of this family---\textsc{MaxSTN} and
\textsc{MinSTN}---are designed to produce orientations with, respectively,
long and short longest paths.  In lack of a formal proof, the authors conjectured, backed by extensive
experimental evidence, that \textsc{MaxSTN} always produces a longest path
at least as long as \textsc{MinSTN}.  The present paper disproves this
conjecture.

%-----------------------------------------------------------------------
\section{\textsc{MaxSTN} and \textsc{MinSTN}}
\label{sec:algorithms}
%-----------------------------------------------------------------------

We briefly recall the algorithms of Papamanthou and
Tollis~\cite{PapamanthouTollis2008}; the reader is referred to that paper
for more details, including proof of correctness and running time. Central to the algorithm is  the notion of a block-cutpoint tree~\cite{HopcroftTarjan73}: Given a connected graph $G$ and a designated sink $t$, a block-cutpoint tree is a tree, rooted $t$ (or at the
unique biconnected component of $G$ containing $t$ if $t$ is not a cut
vertex) that shows the relations among the biconnected components of $G$.  A leaf block is a biconnected component that is a leaf of the block-cutpoint tree.  Every leaf block contains at most one cut vertex of $G$ and its interior contains all nodes in the leaf block except for its cut vertex.

  At any stage of the STN algorithm (Algorithm~\ref{alg-STN}), a node $v^* \neq t$ is eligible for removal
  if $v^*$ belongs to the interior of some leaf block of the current
  block-cutpoint tree of $G_i$, where $G_i$ is the residual graph
  after the first $i$ removals. Removing an eligible node never disconnects $t$ from the remaining graph,
which preserves the invariant needed to produce a valid $st$-orientation.  The STN algorithm (Algorithm~\ref{alg-STN}) maintains the current residual graph $G_i$, a priority queue $Q$ of discovered nodes and the set of currently eligible nodes $E$.

STN uses timestamps to decide the order of removal. Timestamps are initialized to zero for the source $s$.  When a node $u$ is
removed at timer step $\tau$, the timer increments and the timestamp of every
unprocessed neighbor $v$ of $u$ is overwritten with the current timer
value. This overwrite rule (not a max-relaxation) is the key mechanism by which
recently discovered nodes receive high timestamps.  After all nodes except $t$
have been processed, the longest-path length $\ell$ is computed by
the standard DAG relaxation
\begin{equation}
  \ell(v) = \max\bigl\{\,\ell(u) + 1 \;\bigm|\; u \text{ processed before } v,\;
  (u,v) \in E \bigr\} \,, 
  \label{eq:relaxation}
\end{equation}
where $\ell(s) = 0$.
%\begin{figure}
\begin{algorithm}[t]\label{alg-STN}
  \caption{STN$(G, s, t, \text{mode}\in \{\max, \min\}$)}
  \label{alg:stn}
  \begin{algorithmic}[1]
    %\Require Biconnected graph $G$, source $s$, sink $t$,
     %        \textsc{mode} $\in \{\max, \min\}$
    \State $Q \leftarrow \{s\}$;\; $m(s) \leftarrow 0$;\; $\tau \leftarrow 0$;\;
           \textsf{removalOrder} $\leftarrow []$
    \While{$Q \neq \emptyset$}
      \State $E \leftarrow (\text{interior of leaf blocks of block-cutpoint tree of  $G$})\  \cap \ (Q\setminus \{t\})$
      \State $v^* \leftarrow \arg\!\max_{v \in E} m(v)$ \textbf{if} mode $= \max$,
             \textbf{else} $\arg\!\min_{v \in E} m(v)$
             \Comment{ties broken arbitrarily}
      \State Append $v^*$ to \textsf{removalOrder};\;
             $Q \leftarrow Q \setminus \{v^*\}$;\; $\tau \leftarrow \tau + 1$
      \For{each neighbor $u\ne t$ of $v^*$ in $G$}
        \State Add $u$ to $Q$ if not present;\; $m(u) \leftarrow \tau$
               \Comment{overwrite\ $m(u)$ if $u$ is present}
      \EndFor
      \State $G \leftarrow G \setminus \{v^*\}$ \Comment{remove adjacent edges of $v^{*}$ from $G$ as well}
    \EndWhile
    \State \Return longest path length $\ell$ computed by Relaxation~\ref{eq:relaxation}
           on \textsf{removalOrder}
  \end{algorithmic}
\end{algorithm}
%\end{figure}
\textsc{MaxSTN} runs Algorithm~\ref{alg:stn} with mode $= \max$ and
\textsc{MinSTN}  with mode $= \min$.  

\section{Conjecture}

It was observed experimentally (via experiments on planar and non-planar graphs of up to 5,000 nodes) that \textsc{MaxSTN} always
produced an orientation with a longest path at least as long as
\textsc{MinSTN}. No formal proof for this was derived, and the following conjecture was put forth in~\cite{Papamanthou05}.

\begin{conjecture}[Conjecture 3.4.2~\cite{Papamanthou05}]
  \label{conj:main}
  Let $G$ be a biconnected graph with source $s$ and sink $t$.  Let
  $\ell_{\max}$ and $\ell_{\min}$ denote the longest-path lengths
  produced by \textsc{MaxSTN} and \textsc{MinSTN} respectively on $(G,s,t)$.
  Then $\ell_{\max} \geq \ell_{\min}$.
\end{conjecture}

\noindent 
This paper establishes that this conjecture is false by exhibiting a \emph{strong} counterexample, a graph $G$, along with a sink $s$ and $t$, for which there is no execution of $\textsc{MaxSTN}$ that can outperform (in terms of longest-path length) some execution of $\textsc{MinSTN}$.

\section{Strong counterexample}
\label{sec:main}
%-----------------------------------------------------------------------

Our strong counterexample is a biconnected graph $G$ with $n=9$ vertices
and $m=10$ edges---see Figure~\ref{fig:graph}.  In particular, $G$'s vertex
set is $\{0,1,2,3,4,5,6,7,8\}$ and its edge set is
\[
  E = \{(0,4),(0,5),(1,2),(1,5),(2,4),(3,5),(3,7),(4,6),(6,8),(7,8)\}\,.
\]
For the execution of the algorithms we set the source to be $s=1$ and
the sink to be $t=3$.  

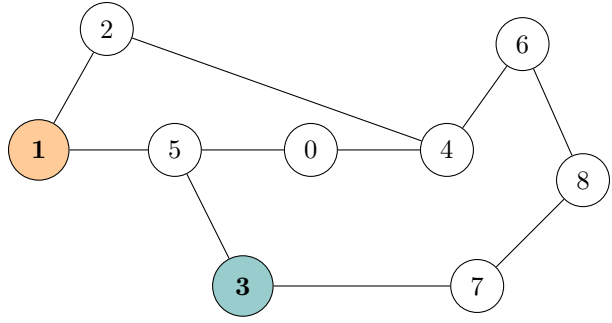
\begin{figure}[t]
\centering
\begin{tikzpicture}[
  every node/.style={circle, draw, minimum size=7mm, font=\small},
  source/.style={circle, draw, fill=orange!40, minimum size=8mm,
                 font=\small\bfseries},
  sink/.style={circle, draw, fill=teal!40, minimum size=8mm,
               font=\small\bfseries},
  every edge/.style={draw, thick}
]
%% Positions: spine 1-5-0-4 runs left to right at y=0;
%%            2 above (top-left), 3 below-centre (sink),
%%            6 top-right, 7 bottom-right, 8 far right.
\node[source] (1) at (0.0,  0.0) {1};
\node         (5) at (1.8,  0.0) {5};
\node         (0) at (3.6,  0.0) {0};
\node         (4) at (5.4,  0.0) {4};
\node         (2) at (0.9,  1.6) {2};
\node[sink]   (3) at (2.7, -1.8) {3};
\node         (6) at (6.4,  1.4) {6};
\node         (7) at (5.8, -1.8) {7};
\node         (8) at (7.2, -0.4) {8};

%% Edges
\draw (1)--(5);
\draw (1)--(2);
\draw (5)--(0);
\draw (5)--(3);
\draw (0)--(4);
\draw (2)--(4);
\draw (4)--(6);
\draw (3)--(7);
\draw (6)--(8);
\draw (7)--(8);
\end{tikzpicture}
\caption{Our strong counterexample.}
\label{fig:graph}
\end{figure}

\begin{figure}[t]
\centering
\begin{tikzpicture}[
  every node/.style={circle, draw, minimum size=7mm, font=\small},
  source/.style={circle, draw, fill=orange!40, minimum size=8mm,
                 font=\small\bfseries},
  sink/.style={circle, draw, fill=teal!40, minimum size=8mm,
               font=\small\bfseries},
  every edge/.style={draw, thick}
]
%% Positions: spine 1-5-0-4 runs left to right at y=0;
%%            2 above (top-left), 3 below-centre (sink),
%%            6 top-right, 7 bottom-right, 8 far right.
\node[source] (1) at (0.0,  0.0) {1};
\node         (5) at (1.8,  0.0) {5};
\node         (0) at (3.6,  0.0) {0};
\node         (4) at (5.4,  0.0) {4};
\node         (2) at (0.9,  1.6) {2};
\node[sink]   (3) at (2.7, -1.8) {3};
\node         (6) at (6.4,  1.4) {6};
\node         (7) at (5.8, -1.8) {7};
\node         (8) at (7.2, -0.4) {8};

%% Edges
\draw [->] (1)--(5);
\draw [->] (1)--(2);
\draw [->] (5)--(0);
\draw [->] (5)--(3);
\draw [->] (0)--(4);
\draw [->] (2)--(4);
\draw [->] (4)--(6);
\draw [<-](3)--(7);
\draw [->] (6)--(8);
\draw [<-] (7)--(8);
\end{tikzpicture}
\caption{\textsc{MinSTN} orientation. }
\label{fig:graph}
\end{figure}
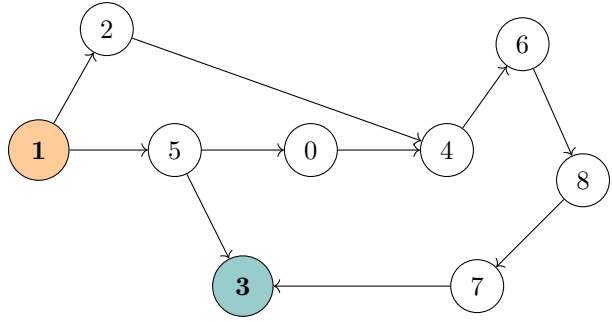

\begin{figure}[t]
\centering
\begin{tikzpicture}[
  every node/.style={circle, draw, minimum size=7mm, font=\small},
  source/.style={circle, draw, fill=orange!40, minimum size=8mm,
                 font=\small\bfseries},
  sink/.style={circle, draw, fill=teal!40, minimum size=8mm,
               font=\small\bfseries},
  every edge/.style={draw, thick}
]
%% Positions: spine 1-5-0-4 runs left to right at y=0;
%%            2 above (top-left), 3 below-centre (sink),
%%            6 top-right, 7 bottom-right, 8 far right.
\node[source] (1) at (0.0,  0.0) {1};
\node         (5) at (1.8,  0.0) {5};
\node         (0) at (3.6,  0.0) {0};
\node         (4) at (5.4,  0.0) {4};
\node         (2) at (0.9,  1.6) {2};
\node[sink]   (3) at (2.7, -1.8) {3};
\node         (6) at (6.4,  1.4) {6};
\node         (7) at (5.8, -1.8) {7};
\node         (8) at (7.2, -0.4) {8};

%% Edges
\draw [->] (1)--(5);
\draw [->] (1)--(2);
\draw [<-](5)--(0);
\draw [->] (5)--(3);
\draw [<-](0)--(4);
\draw [->](2)--(4);
\draw [->](4)--(6);
\draw [<-](3)--(7);
\draw [->](6)--(8);
\draw [<-](7)--(8);
\end{tikzpicture}
\caption{\textsc{MaxSTN} orientation.}
\label{fig:graph}
\end{figure}
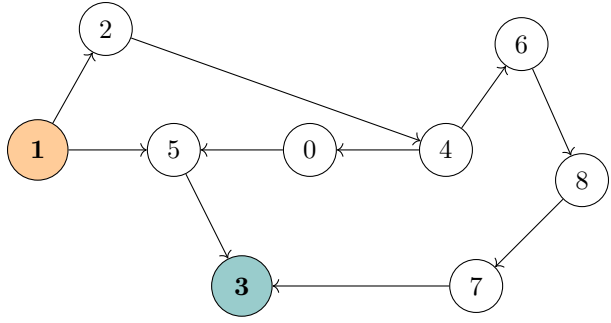

\paragraph{Common execution of \textsc{MaxSTN} and \textsc{MinSTN}.}
Both algorithms must remove node~$1$ first (it is the source).  After
removing~$1$, the block-cutpoint tree of $G\setminus\{1\}$ has exactly
one non-$t$ leaf block, namely $\{2,4\}$, whose sole interior node is~$2$.
Hence both algorithms must remove~$2$ next, irrespective of timestamp or
mode.  After removing~$2$, the residual graph is biconnected with eligible
set $\{4,5\}$.  At this point the timestamps are $m(4)=2$ and $m(5)=1$,
and the two algorithms diverge: \textsc{MaxSTN} picks node~$4$ (higher
timestamp) while \textsc{MinSTN} picks node~$5$ (lower timestamp).
No ties arise at any step of \textsc{MinSTN} execution. \textsc{MaxSTN} can continue the execution with either 0 or 6, but both choices retult into the same orientation.

\paragraph{The execution of \textsc{MaxSTN}.}
There is exactly one execution of \textsc{MaxSTN} on $(G,1,3)$:
\[
  \textsf{removalOrder} = [1,\,2,\,4,\,0,\,5,\,6,\,8,\,7]\,.
\]
The $\ell$-values for this execution are shown in
Table~\ref{tab:max_trace}.  By removing node~$4$ early (step~2),
\textsc{MaxSTN} simultaneously discovers nodes~$0$ and~$6$ with equal
timestamps, which prevents the long sequential chain
$5\to 0\to 4\to 6\to 8\to 7$ from being formed.  The longest
path reaching~$t$ is accordingly capped at $\ell_{\max}=6$.

\begin{table}[t]
\centering
\caption{\textsc{MaxSTN} execution with removal order
         $[1,2,4,0,5,6,8,7]$, giving $\ell_{\max}=6$.}
\label{tab:max_trace}
\begin{tabular}{cccr}
\toprule
Step & Removed node $v$ & $m(v)$ at removal & $\ell(v)$ \\
\midrule
0 & 1 & 0 & 0 \\
1 & 2 & 1 & 1 \\
2 & 4 & 2 & 2 \\
3 & 0 & 3 & 3 \\
4 & 5 & 4 & 4 \\
5 & 6 & 3 & 3 \\
6 & 8 & 6 & 4 \\
7 & 7 & 7 & 5 \\
\midrule
$t=3$ & --- & --- & $\mathbf{6}$ \\
\bottomrule
\end{tabular}
\end{table}

\paragraph{The execution of \textsc{MinSTN}.}
There is exactly one execution of \textsc{MinSTN} on $(G,1,3)$:
\[
  \textsf{removalOrder} = [1,\,2,\,5,\,0,\,4,\,6,\,8,\,7]\,.
\]
The $\ell$-values for this execution are shown in
Table~\ref{tab:min_trace}.  By choosing node~$5$ at step~2 (lower
timestamp) instead of~$4$, \textsc{MinSTN} preserves the sequential chain
$5\to 0\to 4\to 6\to 8\to 7\to t$: each node in this chain has exactly
one already-processed predecessor at the moment it is removed, so their
$\ell$-values increment one by one.  The resulting longest path has
length~$\ell_{\min}=7$.

\begin{table}[t]
\centering
\caption{\textsc{MinSTN} execution with removal order
         $[1,2,5,0,4,6,8,7]$, giving $\ell_{\min}=7$.}
\label{tab:min_trace}
\begin{tabular}{cccr}
\toprule
Step & Removed node $v$ & $m(v)$ at removal & $\ell(v)$ \\
\midrule
0 & 1 & 0 & 0 \\
1 & 2 & 1 & 1 \\
2 & 5 & 1 & 1 \\
3 & 0 & 3 & 2 \\
4 & 4 & 4 & 3 \\
5 & 6 & 5 & 4 \\
6 & 8 & 6 & 5 \\
7 & 7 & 7 & 6 \\
\midrule
$t=3$ & --- & --- & $\mathbf{7}$ \\
\bottomrule
\end{tabular}
\end{table}

\paragraph{Tiebreak independence.}
Since no ties arise at any step of either execution, the result
$\ell_{\max}=6 < 7=\ell_{\min}$ holds for \emph{every} possible
tiebreaking rule applied to either algorithm.  This constitutes a
strong counterexample to Conjecture~\ref{conj:main}.

%-----------------------------------------------------------------------

\section{Statement on the use of AI}
\label{sec:claude}
%-----------------------------------------------------------------------

The counterexample was developed in an interactive session with
Claude Sonnet 4.6. No code was written by a human. Claude also helped with paper writing. We document our
experience here.

Claude read the paper~\cite{PapamanthouTollis2008} and wrote the Python STN code quickly. Some bugs were spotted and were corrected by further interaction. E.g., cut vertices were included as eligible, the sink block was not excluded, and the timestamp overwrite rule was implemented incorrectly.  

Claude performed exhaustive
tests on $n \leq 6$ but did not perform exhaustive search for $n=9$, claiming it was computationally infeasible, although in the end Claude did suggest another resource that could have made this possible~\cite{nauty-geng-manpage}. During the process, Claude did produce some counterexamples that were confirmed wrong with external validation. Claude also drew some conclusions that we did not verify. E.g., Claude claimed that ``\emph{...exhaustive computer search over all $238$ biconnected graphs on 5 vertices
  ($4{,}760$ ordered pairs $(s,t)$) and all $11{,}368$ biconnected graphs on
  6 vertices ($341{,}040$ ordered pairs) found zero violations...}'', implying that $n=7$ could be the first size where the conjecture does not hold.

%-----------------------------------------------------------------------
\bibliographystyle{plain}

\end{document}